\documentstyle[11pt]{article}
\def\Bbb{\bf}

\begin{document}

\newcommand{\beq}{\begin{equation}}
\newcommand{\eeq}{\end{equation}}
\newcommand{\beqa}{\begin{eqnarray}}
\newcommand{\eeqa}{\end{eqnarray}}

\def\ov{\overline}
\def\onlyif{\rightarrow}
\def\and{\th\th\&\th\th}
\def\iff{\leftrightarrow}
\def\dv{\vdash}
\def\ddv{\dashv\vdash}
\def\openone{\leavevmode\hbox{\small1\kern-3.8pt\normalsize1}}
\def\G{G\"odel}
\def\Schr{Schr\"odinger}
\def\na{na\"\i ve}
\def\maczy{M\c{a}czy\'{n}ski}
\def\sing{\Psi_S}
\def\state{\Psi}
\def\spinvec{{\bf\sigma}}
\def\spina{\sigma_a}
\def\spinb{\sigma_b}
\def\spinc{\sigma_c}
\def\spinx{\sigma_x}
\def\spiny{\sigma_y}
\def\spinz{\sigma_z}
\def\unita{{\bf\hat a}}
\def\unitb{{\bf\hat b}}
\def\unitc{{\bf\hat c}}
\def\unitz{{\bf\hat z}}
\def\identity{{\bf I}}
\def\a{\alpha}
\def\b{\beta}
\def\g{\gamma}
\def\r{\rho}
\def\minus{\,-\,}
\def\eks{\bf x}
\def\kay{\bf k}
\def\lorentz{
  \sqrt{1 - \beta^2} }

\def\ket#1{|\,#1\,\rangle}
\def\bra#1{\langle\, #1\,|}
\def\braket#1#2{\langle\, #1\,|\,#2\,\rangle}
\def\proj#1#2{\ket{#1}\bra{#2}}
\def\expect#1{\langle\, #1\, \rangle}
\def\trialexpect#1{\expect#1_{\rm trial}}
\def\ensemblexpect#1{\expect#1_{\rm ensemble}}
\def\kpsi{\ket{\psi}}
\def\kphi{\ket{\phi}}
\def\bpsi{\bra{\psi}}
\def\bphi{\bra{\phi}}

\def\ditto{\rule[0.5ex]{2cm}{.4pt}\enspace}
\def\th{\thinspace}
\def\ni{\noindent}
\def\thirty{\hbox to \hsize{\hfill\rule[5pt]{2.5cm}{0.5pt}\hfill}}

\def\half{\frac{1}{2}}
\def\third{\frac{1}{3}}
\def\squarthtwo{\frac{1}{\sqrt 2}}
\def\squarth#1{\frac{1}{\sqrt{#1}}}
\def\cubth#1{\frac{1}{^3\sqrt{#1}}}

\def\set#1{\{ #1\}}
\def\setbuilder#1#2{\{ #1:\; #2\}}
\def\Prob#1{{\rm Prob}(#1)}
\def\pair#1#2{\langle #1,#2\rangle}
\def\Id{\bf 1}

\def\dee#1#2{\frac{\partial #1}{\partial #2}}
\def\deetwo#1#2{\frac{\partial\,^2 #1}{\partial #2^2}}
\def\deethree#1#2{\frac{\partial\,^3 #1}{\partial #2^3}}
\newcommand{\xx}{{\scriptstyle -}\hspace{-.5pt}x}
\newcommand{\yy}{{\scriptstyle -}\hspace{-.5pt}y}
\newcommand{\zz}{{\scriptstyle -}\hspace{-.5pt}z}
\newcommand{\kk}{{\scriptstyle -}\hspace{-.5pt}k}
\newcommand{\sx}{{\scriptscriptstyle -}\hspace{-.5pt}x}
\newcommand{\sy}{{\scriptscriptstyle -}\hspace{-.5pt}y}
\newcommand{\sz}{{\scriptscriptstyle -}\hspace{-.5pt}z}
\newcommand{\sk}{{\scriptscriptstyle -}\hspace{-.5pt}k}

\def\openone{\leavevmode\hbox{\small1\kern-3.8pt\normalsize1}}
\normalsize
\title{Incoherent and Coherent Eavesdropping in the 6-state Protocol of Quantum Cryptography}
\author{
H. Bechmann-Pasquinucci and N. Gisin \\
\small
{\it Group of Applied Physics, University of Geneva, CH-1211, Geneva 4,
Switzerland}}
\maketitle
\abstract{
All incoherent as well as 2- and 3-qubit coherent eavesdropping strategies on the 6 state protocol 
of quantum cryptography are classified. For a disturbance of $1/6$, the optimal incoherent
eavesdropping strategy reduces to the universal quantum cloning machine. Coherent eavesdropping 
cannot increase Eve's Shannon information, neither on the entire string of bits,
nor on the set of bits received undisturbed by Bob.
However, coherent eavesdropping can increase as well Eve's Renyi information as her probability 
of guessing correctly all bits. The case that Eve delays the measurement of her probe until
after the public discussion on error correction and privacy amplification is also considered.
It is argued that by doing so, Eve gains only a negligibly small additional information.
}

\normalsize
\section{Introduction}
Quantum cryptography, a protocol based
on quantum physics for secret key agreement between two distant parties \cite{BB84}, 
plays two central roles in the field of quantum information processing \cite{PhysWorld}. 
First, it is the most advanced development in the field of quantum information processing 
and it could
be the very first application of quantum mechanics at the individual quanta level. This is
possible because quantum cryptography can be implemented using only 1-qubit technologies,
usually one photon,
contrary to quantum repeaters \cite{Qrepeater}, 
quantum teleportation \cite{Qteleportation} or, more generally, quantum computers \cite{PhysWorld}
that require the coherent processing of tens or even thousands of qubits. Next, besides this quite
practical role, the analysis of various eavesdropping strategies on quantum 
cryptography systems presents very instructive views on the advantages that one can or cannot
expect from coherent processing of several qubits. Indeed, except the trivial intercept-resend
strategy, eavesdropping requires to let one or several auxiliary qubits interact coherently with 
the qubits send by Alice to Bob. For example, the eavesdropping strategy on
the the 4-state protocol (known as the BB84 protocol \cite{BB84}) optimal from Eve's Shannon
information point of view has been found to be intimately related to the Bell inequality (an
inequality that clearly deals with a 2-qubit scenario) \cite{FGGNP,CiracGisin97}.
Similarly, for the 6-state protocol \cite{6state}, optimal eavesdropping is 
related to optimal quantum cloning \cite{GisinMassar97}.

In this article we analyse general eavesdropping strategies on the 6-state protocol. The
symmetry of this protocol simplifies considerably the analysis (compared to the 4-state protocol),
in particular it reduces the number of parameters necessary to describe general strategies. This
motivates our choice to analyse this protocol, although the experimental demonstrations all
use either the 2-state or the 4-state protocol. In the next section, the 6-state protocol
is described. Section \ref{sec:incoh} defines and analyses general incoherent eavesdropping
strategies, various optimization (Shannon and Renyi information, probability of success on
all or some qubits, etc) are presented and the connection to the quantum cloning
machine is made in section \ref{QCM}. Coherent eavesdropping on pairs of qubits is fully
analysed in section \ref{sec:coh2} and similarly for 3 qubits in section \ref{sec:coh3}. 
The case Eve delays her measurement until after the error correction and privacy amplification
phase of the protocol is discussed in section \ref{sec:xor}.
Cumbersome computations and formulas are summarized in appendices.

\section{The 6-state cryptographic protocol}\label{sec:prot}
The 6-state or 3 bases cryptographic scheme is nothing but the well-known
BB84 4-state scheme with an additional basis \cite{6state}. When represented on the Poincar\'e
sphere the BB84 protocol makes use of the four spin-1/2 states
corresponding to $\pm x$ and $\pm y$ directions. In brief summary; Alice
sends one
of the four states to Bob, who measures the qubits he receives in either
the $x$ or the $y$-basis. A priori this gives a probability 1/2 that Alice
and Bob use the same basis. In other words on average Alice and Bob have
to discard half of the qubits even before they can start extracting their
cryptographic key.

In the 6 state protocol the two extra states correspond to $\pm z$, i.e.
the 6
states are $\pm x$, $\pm y$ and $\pm z$ on the Poincar\'e sphere. In this
case Alice
sends a state chosen freely among the 6 and Bob measures either in the
$x$, $y$ or $z$-basis. Here the a prior probability that Alice and Bob use
the same basis is reduced to 1/3, which means that they have
to discard 2/3 of the transmitted qubits before they can extract a
cryptographic key.
However, this scheme does hold an advantage compared to the BB84 protocol
--- higher symmetry. As it will be seen this fact together with the use of
symmetric eavesdropping strategies dramatically reduced the number of free
variables in the problem under investigation. 

\section{Single qubit attacks --- incoherent eavesdropping}\label{sec:incoh}
In the incoherent eavesdropping strategies Eve performs single qubit
attacks, which means that she investigates each qubit sent by Alice
individually. One of Eve's major concerns is not to reveal herself in a
too straightforward manner, for example by introducing different error rates in
the  different bases, she therefore applies a symmetric eavesdropping
strategy which treats all Alice's qubits on an equal footing \cite{symm}.
For each qubit Eve attaches a probe, initially in
the state $\ket{0}$ and let
the system undergo the following unitary evolution:
\begin{eqnarray}
\begin{array}{lcr}
\ket{k}\otimes\ket{0} &\stackrel{\Bbb U}{\longrightarrow}&
\sqrt{F}\ket{k}\otimes\psi_0^k
               + \sqrt{D}\ket{\kk}\otimes\psi_1^{k}\\
\ket{\kk}\otimes\ket{0} &\stackrel{\Bbb U}{\longrightarrow}&
\sqrt{F}\ket{\kk}\otimes{\psi_0}^{\sk}
               + \sqrt{D}\ket{k}\otimes\psi_1^{\sk} \end{array}
\label{Uinc}
\end{eqnarray}
where $k=x,y,z$, $F$ is the fidelity, $D=1-F$ is the disturbance and
$\psi_j^k$ denotes the state of Eve's probe in the case where 
Alice sent the qubit in state $k$ and Bob received it disturbed ($j=1$)
or undisturbed ($j=0$) (i.e. $j$ indicates the number of disturbed qubits). 
Recall that the disturbance
is the probability that Bob gets a wrong bit when he measures the qubit in the bases
compatible with the state send by Alice (sometimes disturbance is called QBER for Quantum
Bit Error Rate). The $\psi_j^k$ are not arbitrary, but determined by the constrain that the
transformation (\ref{Uinc}) is unitary.
Using the Schmidt decomposition \cite{PeresBook} and unitarity of the interaction ${\Bbb U}$ the
${\psi_0}$ and ${\psi_1}$-states can be chosen such that
$\braket{{\psi_0}^{i}}{{\psi_1}^{j}}= 0$ for all $i,j=x,y,z$.
For example, for the $z$-states the unitary evolution result in:
\begin{eqnarray}
\begin{array}{lcr}
\ket{z}\otimes\ket{0} &\stackrel{\Bbb U}{\longrightarrow}&
\sqrt{F}\ket{z}\otimes\psi_0^z
               + \sqrt{D}\ket{\zz}\otimes\psi_1^z\\
\ket{\zz}\otimes\ket{0} &\stackrel{\Bbb U}{\longrightarrow}&
\sqrt{F}\ket{\zz}\otimes\psi_0^{\sz}
               + \sqrt{D}\ket{z}\otimes\psi_1^{\sz} \end{array}
\label{eq:transz}
\end{eqnarray}
The parametrization of $\psi_0^z$ and $\psi_1^z$ can be conveniently and without loss
of generality chosen as
\begin{eqnarray}
\begin{array}{ll}
\psi_0^z=(1,0,0,0)~~~,&~~~
\psi_0^{\sz}=(\cos{a},0,\sin{a},0)\\
\psi_1^z=(0,1,0,0)~~~,&~~~
\psi_1^{\sz}=(0,\cos{b},0,\sin{b})\\
\end{array}
\end{eqnarray}

Note that the 4 dimension of the Hilbert space of Eve's probe is not an assumption, but a
consequence of the fact that only 4 states, $\psi_0^{\pm z}$ and $\psi_1^{\pm z}$, appear in eq.
(\ref{eq:transz}). The $x$-states may be expressed in terms of the $z$-states in the
following way; 
\begin{eqnarray}
\begin{array}{ll}
\ket{x}=\frac{1}{\sqrt{2}}(\ket{z}+\ket{\zz})&
\ket{\xx}=\frac{1}{\sqrt{2}}(-i\ket{z}+i\ket{\zz})\\
\ket{z}=\frac{1}{\sqrt{2}}(\ket{x}+i\ket{\xx})&
\ket{\zz}=\frac{1}{\sqrt{2}}(\ket{x}-i\ket{\xx})\\
\label{eq:xandz}
\end{array}
\end{eqnarray}
When substituting the states in (\ref{eq:transz}) with the ones in
(\ref{eq:xandz}), one finds, using the linearity of ${\Bbb U}$:  
\begin{eqnarray}
\begin{array}{lccl}
\ket{x}\otimes\ket{0} &\stackrel{\Bbb U}{\longrightarrow}& &
\frac{1}{\sqrt{2}}\left( \sqrt{F}\ket{z}\otimes\psi_0^z
               + \sqrt{D}\ket{\zz}\otimes\psi_1^z\right.\\
               & &+& \left.\sqrt{F}\ket{\zz}\otimes\psi_0^{\sz}
               + \sqrt{D}\ket{z}\otimes\psi_1^{\sz}\right)\\
&=& &\sqrt{F}\ket{x}\otimes \frac{1}{2}\left(
\left( \psi_0^z+{\psi}_0^{\zz}\right)+
\frac{\sqrt{F}}{\sqrt{D}}\left(\psi_1^z+\psi_1^{\sz}\right)\right)\\
& &+&  \sqrt{D}\ket{\xx}\otimes \frac{1}{2}\left(
i\left(\psi_0^z-\psi_0^{\sz}\right)-i\frac{\sqrt{F}}{\sqrt{D}}
\left(\psi_1^z+ 
\psi_1^{\sz}\right)\right)
\end{array}
\end{eqnarray}
where the ${\psi}_0^{x}$ and ${\psi}_1^{x}$ are now expressed in terms of
${\psi}_0^{\pm z}$ and ${\psi}_1^{\pm z}$:
\begin{eqnarray}
\begin{array}{c}
{\psi}_0^{x}\equiv \frac{1}{2}\left(
\left( \psi_0^z+\psi_0^{\sz}\right)+
\frac{\sqrt{F}}{\sqrt{D}}\left(\psi_1^z+\psi_1^{\sz}\right)\right)\\
{\psi}_1^{x}\equiv\frac{i}{2}\left(
\left(\psi_0^z-\psi_0^{\sz}\right)-\frac{\sqrt{F}}{\sqrt{D}}
\left(\psi_1^z-
\psi_1^{\sz}\right)\right)
\end{array}
\end{eqnarray}
In a similar way the states ${\psi}_0^{\sx}$ and ${\psi}_1^{\sx}$ can be
found.
When requiring that $\braket{{\psi}_0^{x}}{{\psi}_0^{x}}=1$ one obtains the
following expression for the fidelity 
\begin{eqnarray}
F=\frac{1+\cos{b}}{2-\cos{a}+\cos{b}}
\end{eqnarray}
Going through exactly the same procedure for the $y$-states and again
requiring that $\braket{{\psi}_0^{y}}{{\psi}_0^{y}}=1$ one obtains in this
case the following expression for the fidelity
\begin{eqnarray}
F=\frac{1-\cos{b}}{2-\cos{a}-\cos{b}}
\label{Fincoh}
\end{eqnarray}
Since the fidelity has to be the same in the two bases, the constrain 
$\cos{b}=0$ is imposed, which means that the
fidelity is given by,
\begin{eqnarray}
F=\frac{1}{2-\cos{a}}
\label{Fincoh}
\end{eqnarray}
Notice that this implies that 
\begin{eqnarray}
\psi_1^z=(0,1,0,0)~~~,~~~
\psi_1^{\sz}=(0,0,0,1)
\end{eqnarray}
in other words, we now have that $\braket{\psi_1^z}{\psi_1^{\sz}}=0$, which implies that
Eve can gain full information about the qubits received disturbed by Bob.
Let us again emphasize that the $x-y$ symmetry assumption which guaranties (\ref{Fincoh}) is not
a restriction, since all eavesdropping strategies are equivalent to a symmetric one 
\cite{FGGNP,CiracGisin97}. Notice furthermore that incoherent eavesdropping strategies on the 6-state 
protocol can be parameterized by a single real parameter $a$, in opposition to the 
4-state case which requires 2 real parameters.

Eve, of course, waits to hear which basis was used by Alice before she
performs her measurements. For symmetry reasons it is therefore enough to
consider the case where Alice sent a qubit in the $z$-basis; Eve first has
to distinguish between the $\psi_0$ and the $\psi_1$ states. Since these
states are orthogonal they can be separated perfectly with a standard von
Neumann measurement --- as already mentioned
this means that Eve knows when Bob has received an error. 
Once she knows whether she has a $\psi_0$ or a
$\psi_1$-state, she has to distinguish between 
$\psi_0^z$ and $\psi_0^{\sz}$ or between $\psi_1^z$ and $\psi_1^{\sz}$.
The overlap between $\psi_0^z$ and $\psi_0^{\sz}$ is 
$\braket{\psi_0^z}{\psi_0^{\sz}}=\cos{a}$, which means that Eve can 
make the right identification with probability
$p(s)=\frac{1}{2}(1+\sin{a})$ and make an error with probability
$p(f)=\frac{1}{2}(1-\sin{a})$ \cite{PeresBook}. Eve can distinguish the two   
$\psi_1$-states perfectly since they are orthogonal. 

It is now possible to evaluate Eve's probability of guessing the qubit
correctly, $P(G)$, the probability of guessing the qubit correctly when Bob received it
undisturbed $P(G|undist.)$, Eve' Shannon Information, $I_S$, and her Renyi
Information \cite{RenyiInfo}, $I_R$, in terms of the obtained fidelity and probabilities:
\begin{eqnarray}
P(G)=F\cdot p(s)+D
\label{PG}
\end{eqnarray}
\begin{eqnarray}
P(G|undist.)=p(s)=\frac{1}{2}(1+\sin{a})
\label{PGundist}
\end{eqnarray}
\begin{eqnarray}
I_S=F\cdot S_S(p(f))+D\cdot S_S(1)
\end{eqnarray}
where $S_S(p(f))=1+p(f){\log}_{2}p(f)+p(s){\log}_{2}p(s)$ is Eve's
Shannon information on the ${\psi_0}$-states and $S_S(1)=1$ is her Shannon
information on the ${\psi_1}$-states --- remember that she has full
information on the ${\psi_1}$-states. Finally,
\begin{eqnarray}
I_R=F\cdot S_R(p(f))+D\cdot S_R(1)
\end{eqnarray}
where $S_R(p(f))=1+{\log}_{2}(p^{2}(f)+p^{2}(s))$ is Eve's Renyi
Information on the ${\psi_0}$-states and
$S_R(1)=1$ is her Renyi information on the ${\psi_1}$-states.

To conclude this section let us make two remarks:
\begin{enumerate}
\item According to relation (\ref{Fincoh}) the fidelity $F$ can't decrease below
$1/3$. This implies that there is no way to spin-flip all the 6 states with a probability
of success larger than $2/3$ (which corresponds to the measurement fidelity 
\cite{MassarPopescu95}).
\item Using the Peres-Horodecki separability criterion for 2-qubit states 
\cite{PeresHorodecki} one can prove that Alice and Bob's qubits remain entangled
for $D<1/3$. Hence, using quantum purification and quantum privacy amplification 
\cite{QPA} Alice and Bob can establish a secure secrete key for all $D<1/3$
(note that $D=1/3$ corresponds to the simple intercept-resend strategy).
\end{enumerate}

\subsection{A special case: The Universal Quantum Cloning Machine}\label{QCM}
For increasing disturbance, Bob's fidelity $F$ decreases, while Eve's probability $P(G)$ 
of guessing the bit correctly increases (i.e. Eve's fidelity increases). 
Hence, they cross at a specific value of the fidelity 
F=P(G). This happens for $F=5/6$. For this value of the fidelity the incoherent eavesdropping
strategy turns out to be  precisely identical to the Universal Quantum Cloning Machine introduced
by Bu\v{z}ek and Hillery \cite{Buzek} and proven to be optimal in 
\cite{optimalQCM,GisinMassar97,QCMnoSignal}. This is
seen from the following:
\begin{eqnarray}
\begin{array}{lcl}
{\Bbb
U}_{QCM}\ket{z}&=&\sqrt{\frac{2}{3}}\ket{z,z,z}+\sqrt{\frac{1}{6}}\left(
\ket{z,\zz,\zz}+\ket{\zz,z,\zz}\right)\\
{\Bbb
U}_{QCM}\ket{\zz}&=&\sqrt{\frac{2}{3}}\ket{\zz,\zz,\zz}+
\sqrt{\frac{1}{6}}\left(
\ket{\zz,z,z}+\ket{z,\zz,z}\right)\\
\end{array}
\label{UQCM}
\end{eqnarray}
Hence, the following identifications:
\begin{eqnarray}
\begin{array}{lcl}
\sqrt{F}\psi_0^z&=&\sqrt{\frac{2}{3}}\ket{z,z}+\sqrt{\frac{1}{6}}
\ket{\zz,\zz}\\
\sqrt{D}\psi_1^z&=&\sqrt{\frac{1}{6}}\ket{z,\zz}\\
\sqrt{F}\psi_0^{\sz}&=&\sqrt{\frac{2}{3}}\ket{\zz,\zz}+\sqrt{\frac{1}{6}}
\ket{z,z}\\
\sqrt{D}\psi_1^{\sz}&=&\sqrt{\frac{1}{6}}\ket{\zz,z}\\
\end{array}
\end{eqnarray}
lead to $F=5/6$, $D=1/6$ and that
$\braket{\psi_0^z}{\psi_0^{\sz}}=\cos{a}=4/5$ and
$\braket{\psi_1^z}{\psi_1^{\sz}}=0$. 

The states on the right hand side of eq. (\ref{UQCM}) belong to the 3 qubit Hilbert space.
The first qubit is the original one, the second one is the copy and the third one the internal
(2-dimensional) state of the cloning machine. In quantum cloning the picture is that the
original and the clone correspond to the output. In this case the internal state of the 
cloning machine provides no information at all neither on the original state nor on the
success of the cloning process \cite{GisinMassar97}. In the case of eavesdropping, on the
opposite, the original qubit is viewed as the output of the eavesdropping machine, while
Eve keeps the clone and the machine. Next, she waits to know the basis used by Alice. 
Using then a coherent measurement on both the clone and the 
copy-machine, Eve can determine whether or not the original qubit going to Bob is disturbed
(the measurement has to be coherent because the subspaces spanned by the $\psi_0^{\pm z}$ and
by the $\psi_1^{\pm z}$ are entangled).
Moreover, in case Bob's qubit is disturbed, Eve can gain full information on the original
qubit, contrary to the quantum cloning situation.

This connection between incoherent eavesdropping on the 6-state protocol and the universal quantum
cloning machine replaces the connection that was found between incoherent
eavesdropping on the 4-state protocol and the Bell-CHSH inequality 
\cite{CHSH,GisinHuttner97,FGGNP,CiracGisin97}. This could be
expected \cite{FuchsGuess}, since the Bell-CHSH inequality involves 4 states that lie in a
plane on the Poincar\'e sphere similarly to the 4-state protocol, 
while the universal quantum cloning machine treats all states
symmetrically, similarly to the 6-state protocol. The main difference between the two
connections is that for the 4-state protocol it happens when Eve and Bob Shannon
information are equal, while for the 6-state protocol it happens when Eve and Bob fidelities
(ie mean probability of correct guess) coincide. This can be traced back to the fact that
the Bell inequality is a characteristic of mutual information between two random variables
\cite{BellInfo} while the universal quantum cloning is, by definition, optimized from the fidelity
point of view. This suggest to look at the eavesdropping strategy for the 6-state protocol
when Eve and Bob information coincide, but we could not find any interesting connection there.

\section{2-qubit Coherent eavesdropping}\label{sec:coh2}
In the case of 2-qubit coherent eavesdropping Eve attaches one probe to two of the qubits
sent by Alice. In order to avoid easy detection Eve makes her
eavesdropping on qubits which are not necessarily sent successively. Again, after
attaching her probe to the qubits, she let the system undergo a unitary
transformation, which in this case looks as follows ($k,k'=x,y,z$):
\begin{eqnarray}
\begin{array}{lcclcl}
\ket{k,k'}\otimes\ket{0}&\stackrel{\Bbb U}{\longrightarrow}
& &\sqrt{\alpha}~\ket{k k'}\otimes\psi_0^{kk'}
&+&\sqrt{\beta}~\ket{k\kk'}\otimes\psi_{12}^{kk'}\\
& &+&\sqrt{\beta}~\ket{\kk k'}\otimes\psi_{11}^{kk'}
&+&\sqrt{\gamma}~\ket{\kk\kk'}\otimes\psi_2^{kk'}\\

\ket{k,\kk'}\otimes\ket{0}&\stackrel{\Bbb U}{\longrightarrow}
& &\sqrt{\beta}~\ket{k k'}\otimes\psi_{12}^{k\sk'}
&+&\sqrt{\alpha}~\ket{k\kk'}\otimes\psi_0^{k\sk'}\\
& &+&\sqrt{\gamma}~\ket{\kk k'}\otimes\psi_2^{k\sk'}
&+&\sqrt{\beta}~\ket{\kk\kk'}\otimes\psi_{11}^{k\sk'}\\

\ket{\kk,k'}\otimes\ket{0}&\stackrel{\Bbb U}{\longrightarrow}
& &\sqrt{\beta}~\ket{k k'}\otimes\psi_{11}^{\sk k'}
&+&\sqrt{\gamma}~\ket{k\kk'}\otimes\psi_2^{\sk k'}\\
& &+&\sqrt{\alpha}~\ket{\kk k'}\otimes\psi_0^{\sk k'}
&+&\sqrt{\beta}~\ket{\kk\kk'}\otimes\psi_{12}^{\sk k'}\\

\ket{\kk,\kk'}\otimes\ket{0}&\stackrel{\Bbb U}{\longrightarrow}
& &\sqrt{\gamma}~\ket{k k'}\otimes\psi_2^{\sk\sk'}
&+&\sqrt{\beta}~\ket{k\kk'}\otimes\psi_{11}^{\sk\sk'}\\
& &+&\sqrt{\beta}~\ket{\kk k'}\otimes\psi_{12}^{\sk\sk'}
&+&\sqrt{\alpha}~\ket{\kk\kk'}\otimes\psi_0^{\sk\sk'}\\
\end{array}
\label{Ucoh}
\end{eqnarray}
In these notations, the $\psi_0^{i,j}$ denote Eve's probe state in the case Bob receives
both qubits undisturbed. The $\psi_{11}^{i.j}$, $\psi_{12}^{i.j}$ and $\psi_2^{i.j}$ 
correspond to the case
that the first, second and both qubits are disturbed, respectively.
The fidelity $F$ and the disturbance $D$, as determined by Alice and Bob, are given by
\begin{eqnarray}
F=\alpha + \beta ~~~~,~~~~D=\beta+\gamma
\end{eqnarray}
and satisfy 
\begin{eqnarray}
F+D=\alpha +2\beta +\gamma =1. 
\label{ap2bpg}
\end{eqnarray}
As for the incoherent case, we write the states
(\ref{Ucoh}) in the Schmidt decomposition \cite{PeresBook}, hence all sets 
$\{\psi_0^{i,j}$, $\psi_{11}^{i,j}$, $\psi_{12}^{i,j}$, $\psi_2^{i,j}\}$, 
$i,j=\pm x,\pm y,\pm z$ are formed of four mutually orthogonal 
normalized states.  Since we are
considering (without loss of generality \cite{CiracGisin97}) symmetric eavesdropping
strategies this imposes some restrictions on the scalar products which
characterize the unitary operation $\Bbb U$ used in Eve's attack. Here
this means that the scalar products have to be invariant under the
exchange of $+$ and $-$ directions of any of the two qubits, under the
exchange of the state of the first qubit with the second qubit and finally
under the change of basis. 

It is possible to divide the scalar products into 10 different groups (for
details see Appendix A), each group defining a free parameter. When
imposing the symmetry and unitarity conditions the number of independent
parameters is reduced to only two real ones. (The 4-state protocol requires
5 real parameters \cite{CiracGisin97}). These two parameters can be
chosen in several different ways, but a convenient one for the problem
under investigation is to chose $\alpha$, $\beta$ and $\gamma$, while keeping in mind
the normalization condition (\ref{ap2bpg}), and
express the remaining parameters in terms of these three.

This dramatic reduction in parameters has lead to the following conditions
on the states of Eve's probe. First all scalar products between states of different kinds
(e.g. a $\psi_0$ with a $\psi_{11}$) vanishes. Consequently Eve can know whether Bob received both
qubits undisturbed, or both disturbed, or the first one disturbed but not the second one, 
or vice versa. Next, all 4 ${\psi_2}$-states are mutually orthogonal. 
Third, all ${\psi_{11}}$- and ${\psi_{12}}$-states
with different indices for qubit received disturbed by Bob are also orthogonal. Hence, as
for incoherent eavesdropping, Eve can gain full information on all qubits received disturbed
by Bob. Finally, for the
${\psi_{11}}$- and ${\psi_{12}}$-states the only scalar products different from zero are:
\begin{eqnarray}
\braket{\psi_{12}^{zz}}{\psi_{12}^{\sz z}}&=&\braket{\psi_{12}^{\sz\sz}}{\psi_{12}^{z\sz}}= \nonumber \\
\braket{\psi_{11}^{zz}}{\psi_{11}^{z\sz}}&=&\braket{\psi_{11}^{\sz\sz}}{\psi_{11}^{\sz z}}=
\frac{\beta-\gamma}{\beta}
\label{eq:numu}
\end{eqnarray}
and finally non of the ${\psi_0}$-states are orthogonal. 

Eve is now left in the following situation: After her eavesdropping
attack, she first has to distinguish between the four types of states her
probe could be left in, i.e. whether she remains with a ${\psi_0}$, ${\psi_{11}}$,
${\psi_{12}}$ or a ${\psi_2}$-state. Since these four types of states are
orthogonal Eve will after her first measurement know exactly which kind
of the four she posses. Suppose she finds that she remains with a
$\psi_2$-state, since the states within this subset are all orthogonal
Eve can again identify with certainty which one of them she has.

For the $\psi_{11}$ and $\psi_{12}$-states she has to distinguish between states which
are not all orthogonal, but fulfill (\ref{eq:numu}), these states can in
an optimal way be chosen as follows:
\begin{eqnarray}
\begin{array}{lclcl}
\psi_{11}^{zz}&=&\cos\theta \cdot {e}_{1} &+& \sin\theta \cdot {e}_{3}\\
\psi_{11}^{z\sz}&=&\cos\theta \cdot {e}_{2} &+& \sin\theta \cdot {e}_{4}\\
\psi_{11}^{\sz z}&=&\sin\theta \cdot {e}_{1} &+& \cos\theta \cdot {e}_{3}\\
\psi_{11}^{\sz\sz}&=&\sin\theta \cdot {e}_{2} &+& \cos\theta \cdot {e}_{4}\\
\end{array}
\label{psitheta}
\end{eqnarray}
where  $\{e_j\}$ is an ortho-normal basis.
Eve again uses a standard von Neumann measurement with eigenstates $e_j$ to distinguish these
states. If  the outcome of her measurement is the  state ${e}_{1}$ she
interprets
this as if the initial state was $\psi_{11}^{zz}$, where as if she finds the
state ${e}_{2}$ she interpret it as $\psi_{11}^{z\sz}$, etc. In this way her
probability for making the correct conclusion is $P_{12}={\cos}^{2}\theta$,
and probability $P_{11}={\sin}^{2}\theta$ for making an error. 
(We use the notation $P_{ij}$ to denote Eve's probability of guessing correctly $j$
qubit when $i$ of the qubits arrive disturbed to Bob).
Similarly for the $\psi_{12}$-states.

For the $\psi_0$-states the situation is more complicated since non of
the states  are orthogonal. Let us again introduce an ortho-normal basis $\{e_j\}$
and expand the $\psi_0$ states \cite{footnote1}:
\begin{eqnarray}
\begin{array}{lclclclcl}
\psi_0^{zz}&=& a\cdot {e}_{1} &+& b\cdot{e}_{2}&+& b\cdot {e}_{3}
&+&c\cdot
{e}_{4}\\
\psi_0^{z\sz}&=&b\cdot {e}_{1} &+&  a\cdot {e}_{2}&+&  c\cdot {e}_{3} 
&+&b\cdot {e}_{4}\\
\psi_0^{\sz z}&=&b\cdot {e}_{1} &+&  c\cdot {e}_{2}&+&  a\cdot {e}_{3}
&+&b\cdot {e}_{4}\\
\psi_0^{\sz\sz}&=&c\cdot {e}_{1} &+&  b\cdot {e}_{2}&+&  b\cdot {e}_{3}
&+&a\cdot {e}_{4}\\
\end{array}
\label{abc}
\end{eqnarray}
where $a,b,c$ satisfy the normalization $a^2 +2 b^2 +c^2 =1$ and constrains imposed by
the scalar products among the $\psi_0$'s, as described in appendix A. Defining
($k,l,k',l'=\pm z$):
\begin{enumerate}
\item $A_j=\alpha\braket{\psi_0^{kl}}{\psi_0^{k'l'}}$=product among $\psi_0$'s with $j$ differences
between the indices {kl} and {k'l'}, $j=1,2$,
\item $B_1=\beta\braket{\psi_{1j}^{kl}}{\psi_{1j}^{k'l'}}$=product among $\psi_{1j}$'s 
with one difference between the indices {kl} and {k'l'} corresponding to the undisturbed
qubit (e.g. $B_1=\beta\braket{\psi_{11}^{kl}}{\psi_{11}^{k-l}}
=\beta\braket{\psi_{12}^{kl}}{\psi_{12}^{-kl}}$).
\end{enumerate}
we obtained:
\beq
\matrix{A_1&=&\alpha-\beta & & B_1&=&\beta-\gamma \cr 
A_2&=&\alpha-2\beta+\gamma & &  } 
\eeq
while all other scalar product vanish. Notice that as in the incoherent eavesdropping case,
Eve gains full information on all
qubits that arrive disturbed to Bob.

As before, Eve interprets the outcome ${e}_{1}$
of her measurement as if the initial state was $\psi_0^{zz}$, the outcome ${e}_{2}$
as if the initial state was $\psi_0^{z\sz}$ etc. Her probability of
making the right conclusion about the state is thus $P_{02}=a^2$, in this case
she will have both qubit sent by Alice correctly. With probability
$P_{01}=2b^2$ she will get the state wrong with the following consequence:
she will have one of the qubits sent by Alice correct and the other wrong.
There is of course two ways to obtain this; either the first qubit is
correct and the second wrong or the first qubit is wrong and the second
correct. Finally Eve will with probability $P_{00}=c^2$ get the state
wrong, with the consequence that she will draw the wrong conclusion about
both qubits. 

$P_{12}$, $P_{11}$, $P_{02}$, $P_{01}$ and $P_{00}$,  can all be expressed in
terms of the three parameters $\alpha$, $\beta$ and $\gamma$, see Appendix B.

It is now possible to compute Eve's probability of guessing both of the
two qubits correctly, $P^c(G)$, her probability $P^c(G|undist.)$ of guessing one qubit correctly 
when Bob received it undisturbed, as well as her Shannon Information,
$I_S^c$ and her Renyi
Informations, $I_R^c$, in terms of the found probabilities;
\begin{eqnarray}
P^c (G)=\alpha \cdot P_{02}+2\beta\cdot P_{12} +\gamma
\end{eqnarray}
\beq
P^c(G|undist.)=\frac{\alpha(P_{02}+\half P_{01})+\beta P_{11}}{\alpha+\beta}
\eeq
\begin{eqnarray}
\begin{array}{lcl}
I_S^c =\alpha\cdot S_S^c (P_{02},\half P_{01},\half P_{01},P_{00})&+&2\beta\cdot
S_S^c (P_{12}, P_{11},0,0)\\
&+&\gamma\cdot S_S^c (1,0,0,0)\\
\end{array}
\end{eqnarray}
where $S_S^c (p_1,p_2,p_3,p_4)=2+p_1 {\log}_{2}p_1+ p_2 {\log}_{2}p_2 +p_3
{\log}_{2}p_3 +p_4 {\log}_{2}p_4$.
\begin{eqnarray}
\begin{array}{lcl}
I_R^c =\alpha\cdot S_R^c (P_{02},\half P_{01},\half P_{01},P_{00})&+&2\beta\cdot
S_R^c (P_{12}, P_{11},0,0)\\
&+&\gamma\cdot S_R^c (1,0,0,0)\\
\end{array}
\end{eqnarray}
where $S_R^c (p_1,p_2,p_3,p_4)=2+{\log}_{2}(p_1^2+p_2^2 +p_3^2 +p_4^2)$.

A special case of 2-qubit attack is of course a double incoherent attack. It is
straightforward, though cumbersome, to check that this corresponds to $\alpha=F^2$,
$\beta=FD$, $\gamma=D^2$
and that all the above formula reduce then to the corresponding ones in section \ref{sec:incoh},
with all of Eve's probe states factoring, e.g. $\psi_{12}^{i,j}=\psi_0^i\otimes\psi_1^j$.

In order to understand what Eve can gain using coherent instead of incoherent
eavesdropping, we investigated numerically all four quantities $P^c(G)$, $P^c(G|undist.)$,
$I^c_S$ and $I^c_R$ searching each time for the optimal value of $\alpha$ for a
given disturbance $D$ and compared this with the corresponding quantity obtained for
the same disturbance with incoherent eavesdropping. Let us first consider $P^c(G)$,
the probability that Eve guesses the bits correctly. Figure 1 presents the results. The
long curve from $D=0$ to $D=0.5$ corresponds to the incoherent eavesdropping strategy, while
the 4 shorter curves correspond to different coherent eavesdropping strategies. For each of
the latter, the parameter $\alpha$ is fixed at the indicated value (7/8, 3/4, 1/2 and 1/4). 
Since $\alpha$ and the disturbance cannot vary independently over the entire range, the 4 curves
are only plotted for possible values of $D$ around the value that maximizes $P^c(G)$. Note that
for 0 disturbance, Eve has a probability 0.25 of guessing correctly both qubits. More
important, this figure shows that coherent attacks can slightly increase Eve's fidelity
$P^c(G)$. This can more clearly be seen on the inset which displays a zoom of the $\alpha=7/8$
case. 
This result differs from the other probability of interest, namely Eve's probability of
rightly guessing a qubit received undisturbed by Bob, $P^c(G|undist.)$. Indeed, numerical evidence
show that $P^c(G|undist.)$ is maximal precisely for $\alpha=F^2$, that is precisely when the
2-qubit coherent attack reduce to double incoherent attacks.
Consequently, the use of coherent eavesdropping strategies can increase Eve's
probability of guessing correctly 2 qubits. However, if one restricts the probability to
those qubits received undisturbed by Bob, then coherent eavesdropping is of no use to
Eve (at least for 2-qubit eavesdropping).

For Eve's information, the situations are depicted on figure 2 (Shannon information) and
3 (Renyi information). It turns out that Eve Shannon information is not increased by
coherent attacks,
while her Renyi information is (slightly) increased. 
It might seem strange that Eve's fidelity increases without a
corresponding increase in her Shannon information, but this can be understood as indicating
that the probability that Eve guesses correctly both qubits is compensated by a
corresponding increase of
her probability to guess wrongly both qubits.

\section{3-qubit coherent eavesdropping}\label{sec:coh3}
The generalization to 3-qubit coherent eavesdropping is now straightforward. In brief, the
unitary interaction between the 3 qubits and Eve's probe (generalizing (\ref{Ucoh})) reads:
\beqa
\ket{k,k,k}\otimes\ket{0}\stackrel{\Bbb U}{\longrightarrow}
& &\sqrt{\alpha}~\ket{k,k,k}\otimes{\psi}_0^{kkk} \nonumber \\
&+&\sqrt{\beta}~(\ket{k,k,\kk}\otimes\psi_{13}^{kkk}+\ket{k,\kk,k}\otimes\psi_{12}^{kkk}
+\ket{\kk,k,k}\otimes\psi_{11}^{kkk} \nonumber \\
&+&\sqrt{\gamma}(\ket{k,\kk,\kk}\otimes\psi_{21}^{kkk}+\ket{\kk,k,\kk}\otimes\psi_{22}^{kkk}
+\ket{\kk,\kk,k}\otimes\psi_{23}^{kkk} \nonumber \\
&+&\sqrt{\delta}~\ket{\kk,\kk,\kk}\otimes{\psi}_3^{kkk} 
\eeqa
where $\psi_0^{kkk}$ denotes Eve's probe state in case none of the qubits is disturbed,
$\psi_{1j}^{kkk}$ in case the jth qubit and only this one is disturbed, $\psi_{2j}^{kkk}$
in case all qubits are disturbed except the jth one, and $\psi_3^{kkk}$ in case all 3 qubits
are disturbed. The fidelity and disturbance read: $F=\alpha+2\beta+\gamma$ and 
$D=\beta+2\gamma+\delta$, with normalization $F+D=1$. As for the 2-qubit case, unitarity
and symmetry imposes severe restrictions on the scalar products. Using the following notations
($k,l,m,k',l',m'=\pm z$):
\begin{enumerate}
\item $A_j=\alpha\braket{\psi_0^{klm}}{\psi_0^{k'l'm'}}$=product among $\psi_0$'s with $j$ differences
between the indices {klm} and {k'l'm'}, $j=1,2,3$,
\item $B_1=\beta\braket{\psi_{1j}^{klm}}{\psi_{1j}^{k'l'm'}}$=product among $\psi_{1j}$'s 
with one difference between the indices {klm} and {k'l'm'} corresponding to an undisturbed
qubit (e.g. $B_1=\beta\braket{\psi_{11}^{klm}}{\psi_{11}^{kl-m}}$),
\item $B_2=\beta\braket{\psi_{1j}^{klm}}{\psi_{1j}^{k'l'm'}}$=product among $\psi_{1j}$'s 
with two differences between the indices {klm} and {k'l'm'} corresponding to the two undisturbed
qubits (e.g. $B_2=\beta\braket{\psi_{11}^{klm}}{\psi_{11}^{k-l-m}}$),
\item $C_1=\gamma\braket{\psi_{2j}^{klm}}{\psi_{2j}^{k'l'm'}}$=product among $\psi_{2j}$'s 
with one difference between the indices {klm} and {k'l'm'} corresponding the undisturbed
qubit(e.g. $C_1=\gamma\braket{\psi_{21}^{klm}}{\psi_{21}^{-klm}}$),
\end{enumerate}

we obtained, see appendix C:
\beq
\matrix{A_1&=&\alpha-\beta & & B_1&=&\beta-\gamma \cr 
A_2&=&\alpha-2\beta+\gamma & & B_2&=&\beta-2\gamma+\delta \cr 
A_3&=&\alpha-3\beta+3\gamma-\delta & & C_1&=&\gamma-\delta}
\eeq
while all other scalar product vanish. Notice that again Eve gains full information on all
qubits that arrive disturbed to Bob.

Once Eve knows which kind of states she has, an information she can reliably get, as in the
previous cases, she is left with the problem of optimally distinguishing among the states of that
kind. The corresponding probabilities are summarized in appendix C. The main result is that
using 3-qubit coherent eavesdropping does not improve Eve' Shannon information. Nor does it
improve Eve's probability of guessing correctly a bit received undisturbed by Bob:
\beqa
P^c(G|undist.)&=&
\frac{\alpha(P_{03}+2P_{02}+P_{01})+2\beta(P_{13}+P_{12})+\gamma P_{23}}{\alpha+2\beta+\gamma} \\
&=&P(G|undist.)^3
\eeqa
where $P(G|undist.)$ is given by relation (\ref{PGundist}).

However, it improves the
probability that Eve guesses correctly all three qubits:
\beqa
P^c(G)&=&\alpha P_{03} + 3\beta P_{13} + 3\gamma P_{23} + \delta \\
&>& P(G)^3
\eeqa
where $P(G)$ is given by relation (\ref{PG}).  

Contrary to the 2-qubit case, the gain
is not neglectable: for a disturbance of 7\% the gain on Eve's probability $P^c(G)$ of guessing all
3 qubits correctly is increased by 6\%, with respect to incoherent eavesdropping. This
contrasts with the 1.7\% increase shown in figure 1 for the 2-qubit case. Note however, 
that this larger gain is over a lower probability $P(G)^3$.

\section{Error correction and privacy amplification}\label{sec:xor}
In this section we discuss the possibility that Eve keeps her probe until after Alice and Bob
have carried out the error correction and the privacy amplification phase of the protocol,
using the public channel. Eve can then measure her probe taking into
account all the information she got from the public channel, optimizing her information on
the final key. We shall consider only a simply error correction and privacy amplification
protocol. This
protocol is far from optimal (most bits are wasted), 
but showing that Alice and Bob are safe using this simple
protocol, would prove that quantum cryptography can be made secure even on noisy channels. The 
protocol goes as follows. After Alice and Bob have recorded their raw key (qubits send and
received in the same basis), they randomly chose pairs of bits and compute their xor sum.
For error correction, they announce the xor value and keep the first bit if and only if they
agree on the xor value (the second bit is always discarded). For privacy amplification, Alice and
Bob do not announce the xor value, but discard the 2 randomly chosen bits while keeping the
xor sum for a new key with improved privacy \cite{BBMprivacy}. Hence, the problem for Eve is to
measure her probe in such a way as to maximize her likelihood to correctly guess the xor sum
of two given qubits (in known bases). Clearly, Eve learns which qubits are paired only after
her probe has interacted with the qubits, hence the interaction between the qubits and the
probes can still be assumed symmetric and identical for all qubits, 
as described in the previous section.

The problem of finding Eve's optimal measurements is a very difficult one. In this section we
shall merely use the best measurement we have found. Due to the symmetry of the problem we 
believe that these measurements are optimal, though we recognize that this is an unproven
assumption. Also, we shall assume that all qubits received disturbed by Bob are removed during
the error correction phase, hence we shall only consider the qubits that Bob receives 
undisturbed.

Let us first consider the case of incoherent eavesdropping, that is Eve attaches one probe per
qubit. For this case the best measurement for Eve that we could find consists simply in
measuring each probe separately (this is more efficient, in particular, than to measure whether
the two probes are jointly in a singlet or triplet state). Then, 
the probability that Eve guesses correctly the xor sum of the two bits reads:
\beq
P_{xor1}=P_s^2+(1-P_s)^2 = 1 - \half\frac{(1-2D)^2}{(1-D)^2}
\eeq
where $P_s$ is the probability that Eve guesses correctly any undisturbed bit,
see (\ref{PGundist}), and $D$ is the disturbance. Figure 4 displays $P_{xor1}$ in
function of the disturbance (lower curve).

Next, we consider the case of 2-qubit coherent eavesdropping. In general the two qubits 
attached to one probe will not be paired by Alice and Bob. However, for the argument, let us
assume that Eve would like to guess the xor value of the bits corresponding to a single probe.
The optimal measurement we found reads:
\beq
P_{xor2}=P_{00}+P_{02}
\eeq
where $P_{0j}$ is the probability that Eve guesses correctly $j$ of the two bits (assuming
both are undisturbed). Using the equations (\ref{eq:endb}) of appendix B
one obtains: 
\beq
P_{xor2}=\frac{1}{4\alpha}\left(3-\alpha-4D+\sqrt{9\alpha-5+6D}\sqrt{\alpha-1+2D}\right)
\eeq
For $D\geq\frac{1}{3}$,
$P_{xor2}$ reaches the maximal value of 1 for $\alpha=\frac{1-D}{2}$. For 
$D\in [\frac{5-\sqrt{13}}{12}..\frac{1}{3}]$, the maximum is reached for $\alpha=
\frac{5-21D+28D^2-12D^3}{4-6D}$. Finally, for $D\leq\frac{5-\sqrt{13}}{12}$ the maximum is
reached for $\alpha=1-D$. Figure 4 displays the maximum values of $P_{xor2}$ in
function of the disturbance $D$ (upper curve). This clearly shows that Eve gains using
2-qubit eavesdropping instead of incoherent eavesdropping (lower curve) (except for 
$D=\frac{3-\sqrt{3}}{6}$). Accordingly, Eve gains using 2-qubit coherent eavesdropping, provided
she is lucky enough that Alice and Bob chose to pair the two qubits. However, such a lucky
coincidence is practically excluded if the total number of qubits is much larger than two.

In order to increase the chance that Alice and Bob randomly chosen pairs of bits are attached
to the same probe, Eve should use n-qubit attacks with large n. Let us consider the case n=3
(still not very large, but large enough that several pairing are possible). For this case, we
found that Eve's optimal measurement provides her with the following probability of a correct
guess of the xor value:
\beq
P_{xor3}=\left(\alpha(P_{03}+\frac{1}{3}P_{02}+\frac{1}{3}P_{01}+P_{00}) + 
\beta(P_{13}+P_{11})\right)\frac{1}{\alpha+\beta}
\eeq
where the explicit form of the $P_{ij}$ are given in appendix C
eqs. (\ref{eq_fail0}) - (\ref{eq_fail3}). Numerical optimization of
$P_{xor3}$ is shown on figure 4 (middle curve). It shows that while Eve gains more information
using a 3-qubit coherent strategy than an incoherent one, she gains less than using a 2-qubit
strategy. Based on this result, we conjecture that Eve gains less using an n-qubit strategy
than using a m-qubit one, with $m<n$. However, this should be balanced against the fact that
the probability that Alice and Bob chose pairs of bit corresponding to the same probe
increases with n.

The argument concludes now as follows. Assume Eve uses n-qubit coherent eavesdropping. 
If, on the one side, n
is too small, then the probability that the pairs of bits used for error correction and privacy
amplification correspond to the same probe is small. In this case Eve does not gain delaying
her measurement until after privacy amplification. On the other side, if n is too large, then 
Eve's information gain is negligibly larger than if she would use incoherent eavesdropping.
Finally, for incoherent attacks, Eve does not gain anything by delaying her measurement until
the privacy amplification phase of the protocol. Hence, under the plausible assumptions that
the measurements used in this section are indeed optimal, we conclude that Eve's optimal
strategy consists in attaching one probe per qubit (i.e. incoherent eavesdropping) and to
measure her probe as soon as she learns the basis. The only additional information she can then
get is during the error correction public discussion (a purely classical problem) for which 
bounds are known.

\section{Concluding remarks}\label{sec:conre}
Coherent eavesdropping strategies are unpractical with today's technology. However, it is
important for the reliability of quantum cryptography to clarify the question of whether
such strategies could in the future affect its security. Assuming that the results presented
in this article for 2- and 3-qubit coherent strategies can be generalized to arbitrary n-qubit
strategies, one should be optimistic for quantum cryptography. Indeed, neither the Shannon
information, nor Eve's probability of guessing correctly a bit received undisturbed by 
Bob \cite{Pc}
can be increased using coherent attacks. Admittedly, a complete analysis of eavesdropping
strategies should incorporate the possibility that Eve keeps her probes until Alice and Bob
have carried out all the public discussion part of the protocol, including the privacy 
amplification protocol. Such an analysis is sketched in section \ref{sec:xor} where we
present arguments based on the classification of eavesdropping strategies presented in 
the previous sections concluding that Eve could only gain a negligible small
addition information which would not affect the security of the entire protocol.

The connection to optimal quantum cloning machines, described in section \ref{QCM}, 
generalizes to the 6-state protocol the connection to the Bell inequality found for
the 4-state protocol \cite{GisinHuttner97,FGGNP,CiracGisin97}. 
Possible connections between coherent eavesdropping and quantum
cloning of higher dimension systems remains an open problem. There is an interesting 
connection between, on the one hand side, the fact that optimal quantum cloning machine 
never produce more
errors than the number of additional copies \cite{GisinMassar97}, and, on the other side,
that in optimal eavesdropping Eve gains full information on all qubits that arrive 
disturbed to Bob: for all disturbed qubits, Eve knows that her copy is perfect (if not, there
would be more errors than additional copies).

Using the highly symmetric 6-state protocol introduced in section \ref{sec:prot}, one can
parameterize all incoherent eavesdropping strategies by a single real parameter and all
2-qubit coherent strategies with 2 real parameters and all 3-qubit strategies with 3
real parameters. Hence, all these cases can be analyzed
in details. Our results show that using coherent attacks, Eve can improve on some tasks, like
for example increasing her probability of guessing both qubits correctly, but that she can't
gain for some other tasks, like for example guessing correctly a qubit received undisturbed by 
Bob. In addition to its relevance for the study of the security of quantum cryptography
over noisy channels, this result illustrates that some tasks can be improved by using
quantum coherence, while some other tasks cannot.

\section*{Acknowledgement}
Stimulating discussions with Bruno Huttner and Sandu Popescu are acknowledged. 
H.B.-P. is supported by the Danish National Science 
Research Council (grant no. 9601645). This work profited also from support by the Swiss
National Science Foundation and from the ISI workshop on Quantum Computation in Torino, July 1997.

\appendix
\section{2-qubit coherent eavesdropping: Constrains on Eve's probes and reduction of parameters}
Eve applies a symmetric eavesdropping
strategy \cite{symm}. This imposes some constrains on the scalar products. Namely,
that the scalar products have to be invariant under the
exchange of $+$ and $-$ directions of any of the two qubits, under the
exchange of the state of the first qubit with the second qubit and finally
under the change of basis. Define the following parameters:
\begin{eqnarray}
\begin{array}{lclclcl}
{A}_{1}&=&\alpha~\braket{\psi_0^{zz}}{\psi_0^{z\sz}}
       &=&\alpha~\braket{\psi_0^{zz}}{\psi_0^{\sz z}}& &\\
       &=&\alpha~\braket{\psi_0^{\sz\sz}}{\psi_0^{z\sz}}
       &=&\alpha~\braket{\psi_0^{\sz\sz}}{\psi_0^{\sz z}}& &\\
{A}_{2}&=&\alpha~\braket{\psi_0^{zz}}{\psi_0^{\sz\sz}}
       &=&\alpha~\braket{\psi_0^{z\sz}}{\psi_0^{\sz z}}& &\\
{B}_{1}&=&\beta~\braket{\psi_{12}^{zz}}{\psi_{12}^{\sz z}}
       &=&\beta~\braket{\psi_{12}^{z\sz}}{\psi_{12}^{\sz\sz}}& &\\
       &=&\beta~\braket{\psi_{11}^{zz}}{\psi_{11}^{z\sz}}
       &=&\beta~\braket{\psi_{11}^{\sz z}}{\psi_{11}^{\sz\sz}}& &\\
{B}_{2}&=&\beta~\braket{\psi_{12}^{zz}}{\psi_{12}^{z\sz}}
       &=&\beta~\braket{\psi_{12}^{\sz z}}{\psi_{12}^{\sz\sz}}& &\\
       &=&\beta~\braket{\psi_{11}^{zz}}{\psi_{11}^{\sz z}}
       &=&\beta~\braket{\psi_{11}^{z\sz}}{\psi_{11}^{\sz\sz}}& &\\
{B}_{3}&=&\beta~\braket{\psi_{12}^{zz}}{\psi_{12}^{\sz\sz}}
       &=&\beta~\braket{\psi_{12}^{\sz z}}{\psi_{12}^{z\sz}}& &\\
       &=&\beta~\braket{\psi_{11}^{zz}}{\psi_{11}^{\sz\sz}}
       &=&\beta~\braket{\psi_{11}^{z\sz}}{\psi_{11}^{\sz z}}& &\\
{C}_{1}&=&\gamma~\braket{\psi_2^{zz}}{\psi_2^{z\sz}}
       &=&\gamma~\braket{\psi_2^{zz}}{\psi_2^{\sz z}}& &\\
       &=&\gamma~\braket{\psi_2^{z\sz}}{\psi_2^{\sz\sz}}
       &=&\gamma~\braket{\psi_2^{\sz z}}{\psi_2^{\sz\sz}}& &\\
{C}_{2}&=&\gamma~\braket{\psi_2^{zz}}{\psi_2^{\sz\sz}}
       &=&\gamma~\braket{\psi_2^{z\sz}}{\psi_2^{\sz z}}& &\\ 
\end{array}
\end{eqnarray} 
Note that $A_j$ involved the scalar product between two $\psi_0$-states (i.e. states
corresponding to both of Bob's qubits undisturbed) with j different
indices, $j=1,2$. $B_1$, $B_2$ and $B_3$ involve states corresponding to one of Bob's qubit disturbed
(i.e. $\psi_{11}$- or $\psi_{12}$-states) with the same index for the undisturbed qubit, the same index
for the disturbed qubit and different indices for both qubit, respectively. Finally, the
$C_j$, $j=1,2$ involve states corresponding to both qubits disturbed (i.e. $\psi_2$-states)
with j different indices.
The symmetry under exchange of the directions and states of
the two qubits together with the unitarity condition gives that the
parameters have to fulfill the following relations: 
\begin{eqnarray}
\begin{array}{ll} 
\alpha-\beta={A}_{1}+{B}_{2} & \beta-\gamma={B}_{1}+{C}_{1}\\ 
{B}_{2}-{B}_{3}={C}_{1}+{C}_{2} & {A}_{1}-{A}_{2}={B}_{3}+{B}_{1}\\ 
\end{array} 
\end{eqnarray} 
In this way the number of parameters have been reduced from 10 down to 5 real
parameters.  By imposing the invariance under change of basis, the number
of parameters is further reduced to only two independent. They can, of
course, be chosen in many different ways, but a convenient one is choosing
$\alpha$, $\beta$ and $\gamma$ and express all other parameters in terms of these three,
while remembering the normalization condition (\ref{ap2bpg}):
\begin{eqnarray}
\begin{array}{lll}
A_1= \alpha-\beta ¬ &A_2=\alpha-2\beta+\gamma &\\
B_1=\beta-\gamma ¬&B_2=0&  B_3=0\\
C_1= 0 & C_2=0& \\
\end{array}
\label{AjBjCj}
\end{eqnarray}
Note that all scalar products between probe-states corresponding to qubits received disturbed
by Bob vanish. Hence, Eve can gain full information on these qubits.

\section{2-qubit coherent eavesdropping: determinations of the $P_{ij}$ probabilities}
Using the overlaps between the various states (given in appendix A)
together with the expressions of the $\psi_{11}$- and 
$\psi_{12}$-states, see equation (\ref{psitheta}), one finds that
$2\sin\theta\cos\theta = {B}_{1}/\beta$, hence
\beqa
P_{12}&=&\cos^2\theta \nonumber \\
&=&\frac{1}{4}(\sqrt{1+(B_1/\beta)}+\sqrt{1-(B_1/\beta)})^2 \nonumber \\
&=&\frac{1+\sqrt{2\gamma/\beta-\gamma^2 /\beta^2}}{2}
\end{eqnarray}

For the $\psi_0$-states, let us introduce the vectors $\vec\psi\equiv(\psi_0^{zz},\psi_0^{z\zz},
\psi_0^{\zz z},\psi_0^{\zz\zz})$ and $\vec e\equiv(e_1,e_2,e_3,e_4)$ whose entries are elements
of Eve's probe Hilbert space. The $e_j$ are the eigenstates of Eve's measurement, hence
they are mutually orthogonal \cite{footnote1}. The entries
of the matrix $X\equiv\ket{\vec e}\bra{\vec\psi}$ are the probability amplitudes for the various
possible outcomes. The matrix $X$ can be easily computed from:
\beq
X^\dagger X=\ket{\vec\psi}\bra{\vec\psi}=\frac{1}{\alpha}\pmatrix{\alpha & A_1 & A_1 & A_2 \cr A_1 & 
\alpha & A_2 & A_1\cr A_1 & A_2 & \alpha & A_1\cr A_2 & A_1 & A_1 & \alpha} 
\eeq
Hence, one obtains the probabilities that define relation (\ref{abc}):
\beqa
\label{abcequiv1}
a&\equiv&\braket{e_1}{\psi_{z,z}}=\frac{1}{4}\left(
2\sqrt{1-\frac{A_2}{\alpha}}+
\sqrt{1+\frac{A_2}{\alpha}+2\frac{A_1}{\alpha}}+
\sqrt{1+\frac{A_2}{\alpha}-2\frac{A_1}{\alpha}}\right)  \\
b&\equiv&\braket{e_2}{\psi_{z,z}}=\frac{1}{4}\left(
\sqrt{1+\frac{A_2}{\alpha}+2\frac{A_1}{\alpha}}-
\sqrt{1+\frac{A_2}{\alpha}-2\frac{A_1}{\alpha}}\right)  \\
c&\equiv&\braket{e_4}{\psi_{z,z}}=\frac{1}{4}\left(
2\sqrt{1-\frac{A_2}{\alpha}}-
\sqrt{1+\frac{A_2}{\alpha}+2\frac{A_1}{\alpha}}-
\sqrt{1+\frac{A_2}{\alpha}-2\frac{A_1}{\alpha}}\right)
\label{abcequiv2}
\eeqa
Finally, the probabilities $P_{0j}$ that Eve guesses correctly $j$ bits
are 
\begin{eqnarray}
P_{02}=a^2~~~~ P_{01}=2b^2 ~~~~{\rm and}~~~~ P_{00}=c^2.
\label{eq:endb}
\end{eqnarray}

\section{3-qubit eavesdropping: probability for Eve's guesses}
If Eve is left with a $\psi_3^{klm}$ state, then she can know all three qubits (recall that all
the 8 $\psi_3^{klm}$ states are mutually orthogonal).

If Eve is left with a $\psi_{2j}^{klm}$ state, then, using the same technique as in appendix B,
one finds that she has probability $P_{23}=\half(1+\sqrt{1-C_1^2/\gamma^2})$ to guess 
correctly all 3 qubits and $P_{22}=1-P_{23}$ to make one error. 

If Eve is left with a $\psi_{1j}^{klm}$ state, then, using the same technique as in appendix B,
one finds that she has probability $P_{1j}$ to guess correctly j qubits, with j=1,2,3. The
formula are the same as for 2-qubit coherent eavesdropping, see (\ref{abcequiv1})
to (\ref{abcequiv2}), but with the
$A_n$ and $\alpha$ replaced by $B_n$ and $\beta$, respectively: 
$P_{13}=a^2$, $P_{12}=2b^2$ and $p_{11}=c^2$.

If Eve is left with a $\psi_{0}^{klm}$ state, then, using again the technique of appendix B,
one finds that she has:\\
Probability to guess correctly all 3 qubits=
\beqa
P_{03}=\frac{1}{64\alpha}&\big(&\sqrt{\alpha+A_3+3(A_1+A_2)}+3\sqrt{\alpha-A_3+A_1-A_2} \nonumber\\
&+&\sqrt{\alpha-A_3-3(A_1-A_2)}+3\sqrt{\alpha+A_3-A_1-A_2}~\big)^2,
\label{eq_fail0}
\eeqa
Probability to guess correctly 2 of the 3 qubits=
\beqa
P_{02}=\frac{3}{64\alpha}&\big(&\sqrt{\alpha+A_3+3(A_1+A_2)}+\sqrt{\alpha-A_3+A_1-A_2} \nonumber\\
&-&\sqrt{\alpha-A_3-3(A_1-A_2)}-\sqrt{\alpha+A_3-A_1-A_2}~\big)^2,
\eeqa
Probability to guess correctly 1 of the 3 qubits=
\beqa
P_{01}=\frac{3}{64\alpha}&\big(&\sqrt{\alpha+A_3+3(A_1+A_2)}-\sqrt{\alpha-A_3+A_1-A_2} \nonumber\\
&+&\sqrt{\alpha-A_3-3(A_1-A_2)}-\sqrt{\alpha+A_3-A_1-A_2}~\big)^2,
\eeqa
Probability to guess correctly none of the 3 qubits=
\beqa 
P_{00}=\frac{1}{64\alpha}&\big(&\sqrt{\alpha+A_3+3(A_1+A_2)}-3\sqrt{\alpha-A_3+A_1-A_2} \nonumber\\
&-&\sqrt{\alpha-A_3-3(A_1-A_2)}+3\sqrt{\alpha+A_3-A_1-A_2}~\big)^2.
\label{eq_fail3}
\eeqa

\newpage
\section*{Figure Captions}
\begin{enumerate}
\item Probability that Eve guesses correctly two bits out of two in function of the 
disturbance D. The long curve from $D=0$ to $D=0.5$ corresponds to the incoherent 
eavesdropping strategy of section \ref{sec:incoh}, while
the 4 shorter curves correspond to different 2-qubit coherent eavesdropping strategies
described in section \ref{sec:coh2}. For each of
the latter, the parameter $\alpha$ is fixed at the indicated value (7/8, 3/4, 1/2 and 1/4). 
This figure shows that coherent attacks can slightly increase Eve's fidelity
$P^c(G)$. This can more clearly be seen on the inset which displays a zoom of the $\alpha=7/8$
case. 

\item Same as figure 1, but for Eve's Shannon information. Coherent eavesdropping does
not improve Eve's Shannon information.

\item Same as figure 1, but for Eve's Renyi information.  Coherent eavesdropping does
slightly improve Eve's Renyi information.

\item Probability that Eve guesses correctly the xor sum of two bits. The lower curve
corresponds to incoherent eavesdropping. The upper and middle curves corresponds to 2-qubit
and 3-qubit coherent
eavesdropping assuming the two bits correspond to qubits attached to the same probe.

\end{enumerate}


\begin{thebibliography}{99}
\bibitem{symm} There
is no loss of generality by assuming that the eavesdropping strategy is
symmetric, since for all strategies there is a symmetric one that produces the same disturbance
on the quantum channel connecting Alice to Bob
and provides Eve with the same set of probabilities, see \cite{FGGNP,CiracGisin97}.
\bibitem{footnote1} This is only possible if the 4 vectors $\psi_0^{\pm z\pm z}$ span a 4-dimensional
       space, an assumption valid for all efficient eavesdropping strategies.
\bibitem{Pc} This seems to be the most relevant quantity, since all bits received disturbed
by Bob are eliminated during the error correction phase of the protocol \cite{Wolf}.
\end{thebibliography}
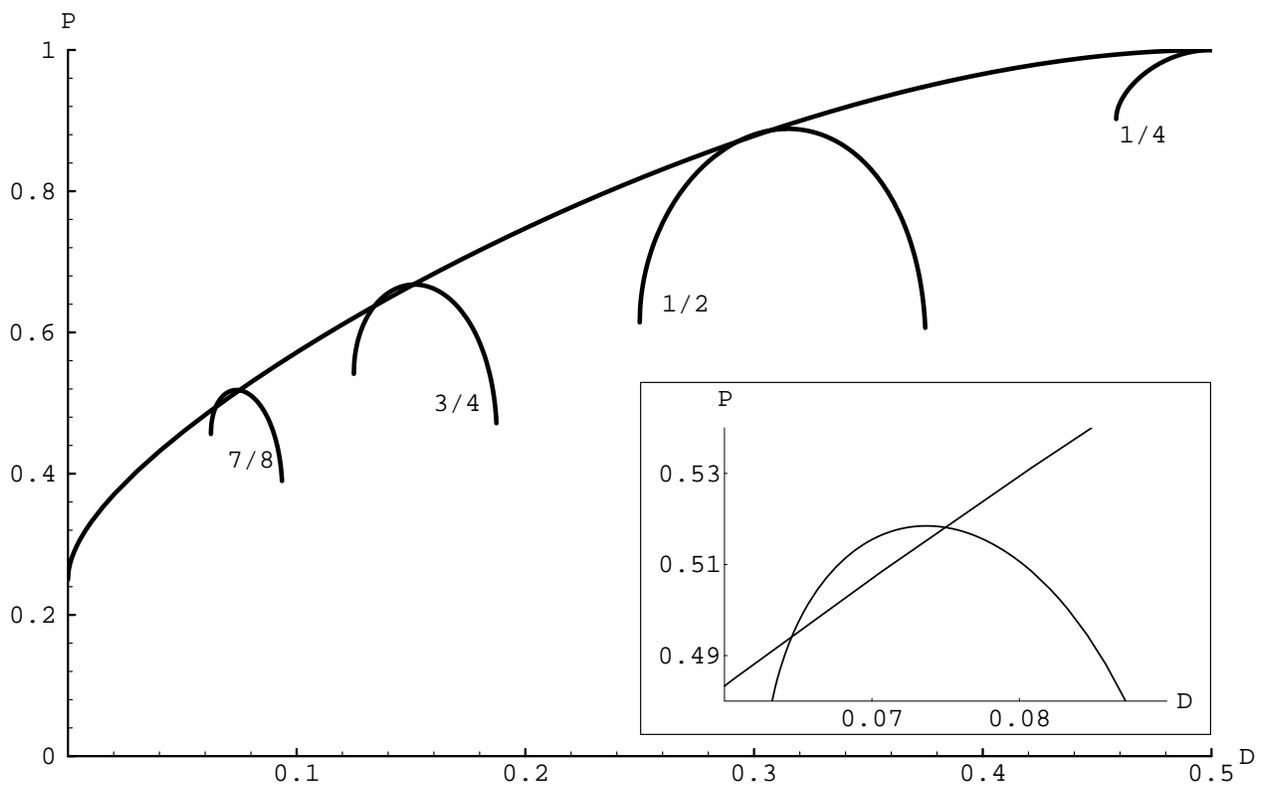
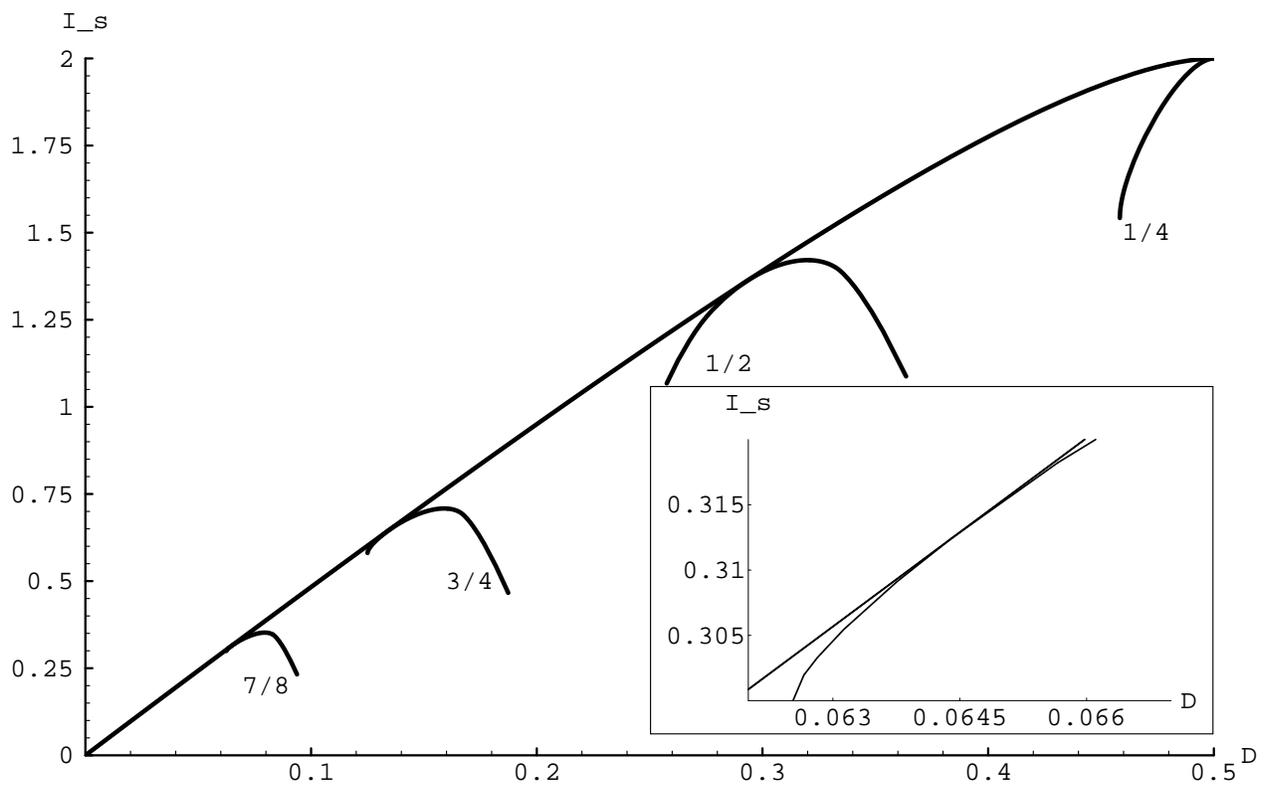
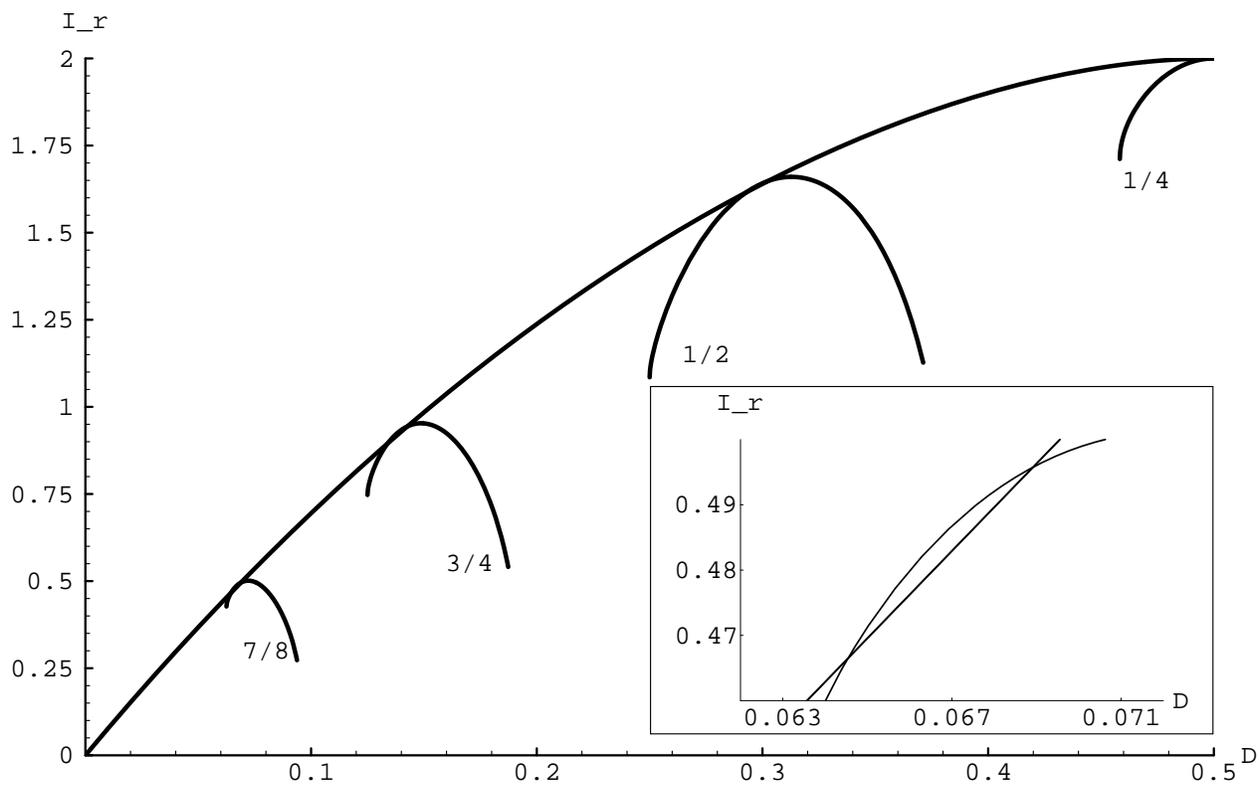
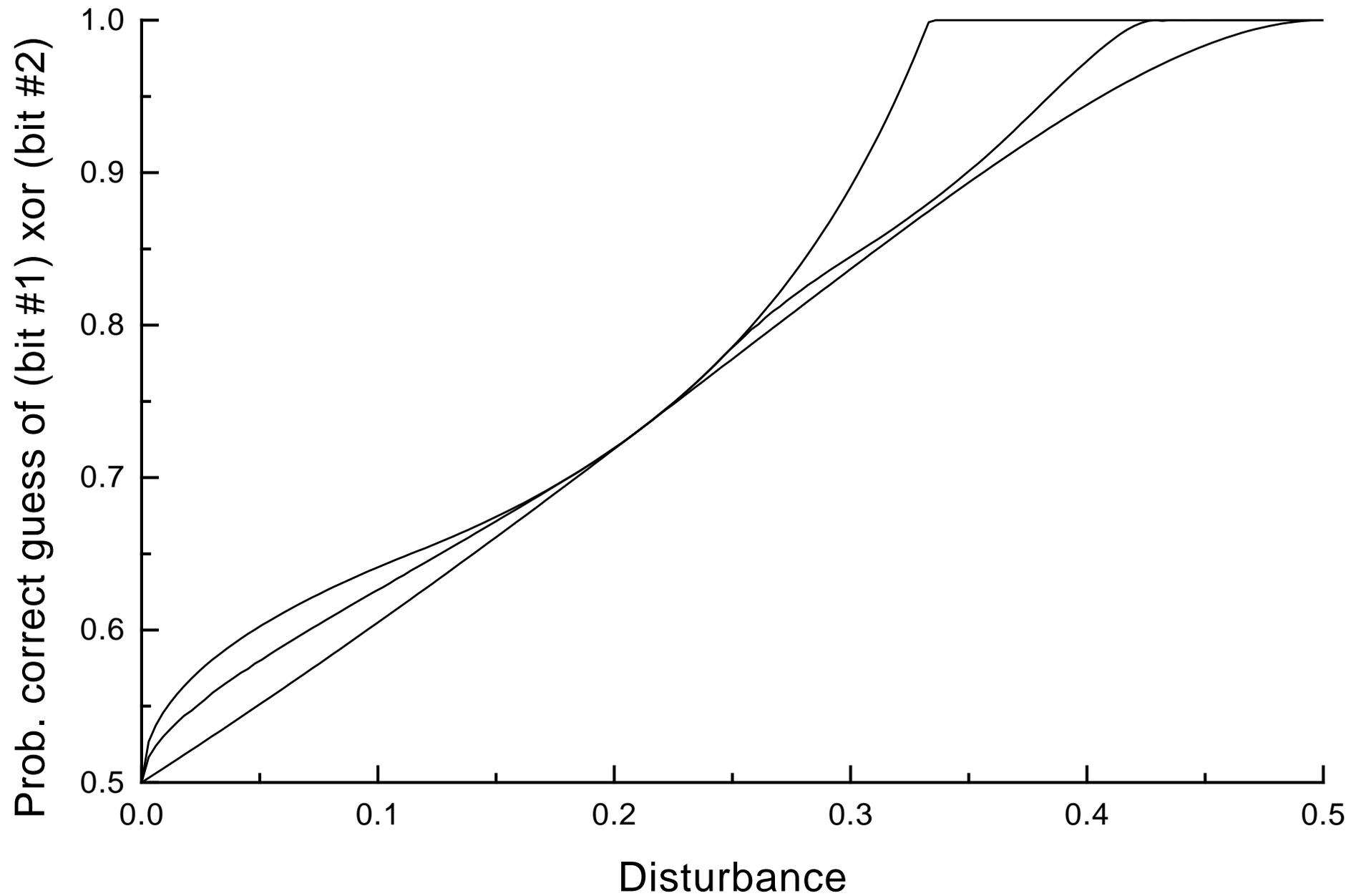
\end{document}